\documentclass{article}

\PassOptionsToPackage{numbers, compress}{natbib}

\usepackage[preprint]{neurips_2021}

\usepackage[utf8]{inputenc} 
\usepackage[T1]{fontenc}    
\usepackage{hyperref}       
\usepackage{url}            
\usepackage{booktabs}       
\usepackage{amsfonts}       
\usepackage{nicefrac}       
\usepackage{microtype}      
\usepackage{graphicx}
\usepackage{subcaption}
\usepackage{amsmath}
\usepackage{multirow}
\usepackage{adjustbox}
\usepackage{bbm}
\usepackage{wrapfig,lipsum}

\usepackage{xcolor}         

\title{Time-Aware Neighbor Sampling for \\Temporal Graph Networks}

\author{
Yiwei Wang$^1$ \ \ \ \ Yujun Cai$^2$ \ \ \ \  Yuxuan Liang$^1$ \ \ \ \  Henghui Ding$^3$ \\ \textbf{Changhu Wang$^3$ \ \ \ \  Bryan Hooi$^1$} \\ 
$^1$ National University of Singapore\\ 
$^2$ Nanyang Technological University \\
$^3$ ByteDance \\ 
\texttt{wangyw\_seu@foxmail.com, \{yujun001,ding0093\}@e.ntu.edu.sg,}\\ 
\texttt{yuxliang@outlook.com, changhu.wang@gmail.com,}\\
\texttt{bhooi@comp.nus.edu.sg}
}

\begin{document}

\maketitle

\begin{abstract}
    We present a new neighbor sampling method on temporal graphs.
    In a temporal graph, predicting different nodes' time-varying properties can require the receptive neighborhood of various temporal scales.
    In this work, we propose the \textbf{TNS} (\textit{Time-aware Neighbor Sampling}) 
    method: TNS learns from temporal information to provide an adaptive receptive neighborhood for every node at any time.
    Learning how to sample neighbors is non-trivial, since the neighbor indices in time order are discrete and not differentiable.
    To address this challenge, we 
    transform neighbor indices from discrete values to continuous ones by interpolating the neighbors' messages.
    TNS can be flexibly incorporated into popular temporal graph networks to improve their effectiveness without increasing their time complexity.
    TNS can be trained in an end-to-end manner.
    It needs no extra supervision and is automatically and implicitly guided to sample the neighbors that are most beneficial for prediction.
    Empirical results on multiple standard datasets show that TNS yields significant gains on edge prediction and node classification.
\end{abstract}

\section{Introduction}
Many real-world graphs are not static but evolving, e.g., edges can appear at any time \cite{rossi2020temporal}.
Nodes may interact either due to gradual trends or fortuitous encounters.
These graphs are called temporal (or dynamic) graphs \cite{xu2020inductive}.
Using static graph methods, e.g., GraphSAGE \cite{hamilton2017inductive}, to model temporal graphs is suboptimal since they cannot capture the evolutionary patterns.
Recently, temporal graph networks (TGNs) \cite{rossi2020temporal,xu2020inductive,kumar2019predicting} have been proposed to support learning on temporal graphs.

\begin{figure}[!tb]
	\centering
	\includegraphics[width=0.7\linewidth]{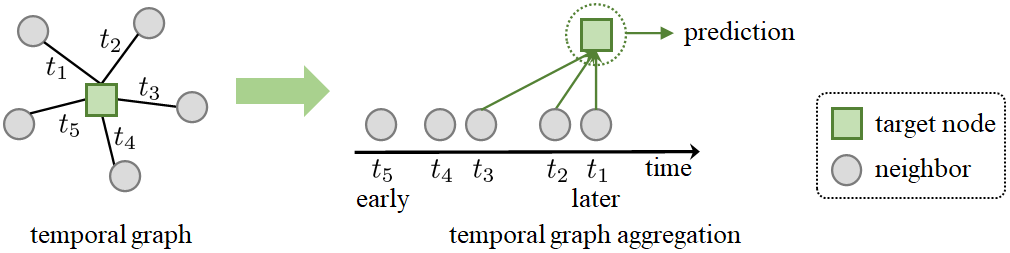}
	\caption{Existing temporal graph aggregation modules \cite{rossi2020temporal} aggregate the most recent neighbors for the target node to obtain its representations and make predictions. \label{fig:agn}}
\end{figure}

Advanced TGNs utilize a temporal graph aggregation module to obtain a target node's embedding \cite{rossi2020temporal,xu2020inductive}, which aggregate the messages from the target node's neighbors.
The embedding is used to predict the target node's properties \cite{trivedi2019dyrep}.
To prevent the number of neighbors from increasing without limitation as time flows, TGNs sample neighbors for the message aggregation (see Fig. \ref{fig:agn}), which improves their efficiency and stabilization \cite{hamilton2017inductive,xu2020inductive,rossi2020temporal}.
Specifically, \cite{hamilton2017inductive} and \cite{xu2020inductive} utilize uniform neighbor sampling, that samples every neighbor with the same probability.
\cite{rossi2020temporal} improves the neighbor sampling by incorporating temporal information, which samples the neighbors that interacted with the target node most recently.

Although sampling the most recent neighbors is generally more effective than uniform sampling \cite{rossi2020temporal}, 
we argue that this may not be the only or the best option to utilize temporal information for neighbor sampling.
First, sampling successive neighbors along the time axis can induce information redundancy.
For example, a student can repeatedly interact with his classmates for several times in a short period because they are interested in a new computer game.
Repeatedly considering such redundant interactions does not provide more useful information.
Second, sampling the most recent neighbors limits the temporal scale of the receptive neighborhood \cite{wang2020nodeaug}, i.e., the neighborhood contributing to the target node's embedding.
Effectively predicting a node's properties may require long-range dependencies.
For example, a woman meets her fitness coach for exercising every Saturday.
Sampling the most recent neighbors cannot capture her long-range exercising trends.

To address the above issues, we propose an expanded neighbor sampling approach, which inserts spacing between the sampled neighbors on the time axis.
We define the expansion rate to control the spacing size. 
With the expansion rate as $r$, every sampled neighbor skips the next $r - 1$ neighbors in time order (see Fig. \ref{fig:exp1}).
The inserted spacing reduces the redundancy between the sampled neighbors, while extending the temporal scale of the receptive neighborhood.

This expanded neighbor sampling fixes a unified expansion rate for all the nodes at any time.
This may be suboptimal since predicting different nodes' time-varying properties can require various expansion rates.
Hence, beyond the unified expansion rate, 
we aim to learn suitable and adaptive expansion rates from temporal information to offer appropriate receptive neighborhoods.
Learning how to sample neighbors is non-trivial, since the neighbor indices in time order are discrete and not differentiable.
In this work, we view the neighbors as image pixels and neighbors' messages as pixel values, and compare the neighbor sampling to the process of image rendering \cite{kirillov2020pointrend}.
Concretely, we transfer the neighbor indices from discrete values to continuous ones by interpolating neighbors' messages, so that the neighbors' messages of any index in time order, even if not an integer, can be accessed.
We encapsulate this idea into a new neighbor sampling method, called \textbf{TNS} (\textit{Time-aware Neighbor Sampling}), which learns expansion rates using an expansion learning module,
and then uses the learned expansion rates to guide the neighbor sampling for message aggregation.

TNS can be incorporated into the popular TGN models to enhance their performance.
It needs no extra supervision and can be trained in an end-to-end style.
We analyze the back-propagation 
on TNS and find that TNS is automatically and implicitly guided to sample the neighbors that are most beneficial for prediction.
In addition, theoretical analysis shows that using our TNS to improve the effectiveness of TGNs does not increase their time complexity.

We evaluate our TNS method on edge prediction and node classification tasks using the standard temporal graph datasets: Reddit \cite{baumgartner2020pushshift}, Wikipedia \cite{wiki}, MOOC \cite{kumar2019predicting}.
We measure its performance 
through the metrics: test accuracy, average precision (AP), and the area under the ROC accuracy curve (AUC), under inductive and transductive settings.
Overall, TNS achieves substantial improvements when applied to popular TGN models \cite{rossi2020temporal,xu2020inductive} 
and enhances them to outperform the baseline methods.

\section{Related Work}
There has been a burst of remarkable work for representation learning on static graphs \cite{perozzi2014deepwalk,grover2016node2vec,kipf2016semi,kipf2016variational,hamilton2017inductive,velivckovic2017graph},
but the work on temporal graphs is much sparser. 
There exist two main classes of temporal graphs: discrete-time dynamic graphs (DTDG) \cite{liben2007link} and continuous-time dynamic graphs (CTDG) \cite{rossi2020temporal}.
DTDG are sequences of static graph snapshots taken at intervals in time, while CTDG can be represented as timed lists of events.
Representation learning on CTDG is more flexible, general and challenging, which is the focus of this paper.

Early models for temporal graph learning focus on DTDGs \cite{dunlavy2011temporal,yu2017link}, which aggregate
graph snapshots and then apply static methods \cite{hisano2018semi,liben2007link,ahmed2016sampling},
or encode each snapshot to produce a series of embeddings \cite{pei2016node,yao2016link}.
More recently, some work consider the CTDGs \cite{kumar2019predicting,trivedi2019dyrep}.
\cite{wang2020apan,xu2020inductive,rossi2020temporal} aggregates the messages from neighbors
through a temporal aggregation module, which present superior performance.
To prevent the number of neighbors from growing without limitation as time flows, \cite{xu2020inductive} uniformly samples the neighbors to improve the efficiency and stabilization.
\cite{rossi2020temporal} improves the neighbor sampling further by incorporating the temporal information, which samples the most recent neighbors.
Our work proposes new approaches for utilizing temporal information for neighbor sampling.
First, our expanded neighbor sampling approach
reduces information redundancy 
and extends
the temporal scale of
receptive neighborhoods by inserting space 
between the sampled neighbors along 
the time axis.
Second, beyond the unified sampling strategy, 
our TNS method provides the adaptive receptive neighborhood for every node at any time in a learnable way.
Our methods improves the effectiveness of popular TGNs without increasing their time complexity.

\section{Methodology}
In this section, we describe our neighbor sampling methods for temporal graph learning.
We first introduce the background and mathematical notations. 
Next, we introduce our expanded neighbor sampling approach
for reducing the information redundancy and extending the time scale of the receptive neighborhood.
In addition, beyond the unified and fixed expanded sampling, 
we propose Time-aware Neighbor Sampling (TNS) 
to provide an adaptive receptive neighborhood for every node at any time.
Finally, we analyze the time complexity of our methods, as well as how our TNS moves the neighbor indices (in time order) in the direction that is most beneficial for prediction.

\subsection{Preliminaries}\label{sec:moti}

In a temporal graph, an interaction between nodes $i$ and $j$ is denoted as a temporal edge $\mathbf{e}_{(i,j)}(t)$
, where $\mathbf{e}$ is the edge attributes \cite{rossi2020temporal}.
We define the \textit{temporal neighborhood} 
of node $i$ at time $t$ as a sequence of temporal neighbors in reverse chronological order
\begin{equation}\label{eq:nei}
\mathcal{T}_i(t) = \big\{\eta(i, t_1), \dots, \eta(i, t_n), \dots, \eta(i, t_{N(i,t)})\big\},
\end{equation}
where the neighbor $\eta(i,t_n)$ corresponds to the temporal edge $\mathbf{e}_{(i, \eta(i,t_n))}(t_n)$, and $t>t_1 \ge\dots\ge t_N \ge 0$ holds, i.e., $\eta(i,t_n)$ is the $n$th most recent temporal neighbor of node $i$ at time $t$.
$N(i,t)$ is the number of all the temporal neighbors of node $i$ before time $t$.
Advanced TGN models
stack $L$ temporal graph aggregation modules to obtain nodes' embeddings to predict nodes' time-varying properties, and finds that $L = 1,2$ generally leads
to the best performance \cite{xu2020inductive,rossi2020temporal}.
Denote the representation of node $i$ from the $l$-th module at time $t$ as $\mathbf{h}_i^{(l)}(t)$.
Existing work \cite{rossi2020temporal} builds 
the message that the $n$th most recent neighbor $\eta(i, t_n)$ conveys to node $i$ at module $l$ as:
\begin{equation}
\mathbf{m}_{i}^{(l)}(n, t) = \mathbf{h}_{\eta(i, t_n)}^{(l-1)}(t)\|\mathbf{e}_{(i,\eta(i, t_n))}\|\phi(t - t_n),
\end{equation}
where $\phi(\cdot)$ represents a generic time encoding \cite{xu2020inductive}, and $\|$ denotes the concatenation operation.
Since $N(i,t)$ can grow without limitations as $t$ increases, TGN models \cite{xu2020inductive,rossi2020temporal} sample the neighbors from $\mathcal{T}_i(t)$ in Eq. \eqref{eq:nei} for message aggregation, which offers efficiency and stabilization.
Recently, \cite{rossi2020temporal} finds that uniformly sampling neighbors is generally less effective than sampling the most recent neighbors, of which the corresponding indices are
\begin{equation}\label{eq:sam}
\mathcal{S}_i(t) = \big\{1,\dots,\min\big(N(i,t),S\big)\big\},
\end{equation}
where $S$ is the number of sampled neighbors. 
Denote the $l$th temporal graph aggregation module as $\mathsf{AGGR}^{(l)}$.
It returns $\mathbf{h}_i^{(l)}(t)$ as:
\begin{equation}\label{eq:agg}
\mathbf{h}_i^{(l)}(t) = \mathsf{AGGR}^{(l)}\big(\mathbf{h}_i^{(l-1)}(t), \{\mathbf{m}_{i}^{(l)}(n, t)\ |\ n \in \mathcal{S}_i(t)\}\big),
\end{equation}
where $n$ refers to the $n$th most recent neighbor (see Fig. \ref{fig:agn}).
$\mathsf{AGGR}^{(l)}$ can be implemented as graph attention, sum, or mean aggregation.
For example, the latter implementation can be:
\begin{align}
\hat{\mathbf{h}}_i^{(l)}(t) &= \mathsf{ReLU}\Big(\mathbf{W}_1^{(l)}\cdot\mathsf{MEAN}\big(\{\mathbf{m}_n^{(l)}(i, t) | n \in \mathcal{S}_i(t) \}\big) + \mathbf{b}_1^{(l)}\Big) \nonumber\\
\mathbf{h}_i^{(l)}(t) &= \mathbf{W}_2^{(l)}\Big(\mathbf{h}_i^{(l-1)}(t)\|\hat{\mathbf{h}}_i^{(l)}(t)\Big) + \mathbf{b}_2^{(l)}, \label{eq:lay}
\end{align}
where $\mathbf{W}_1^{(l)}$, $\mathbf{W}_2^{(l)}$ and $\mathbf{b}_1^{(l)},\mathbf{b}_2^{(l)}$ are the learnable weights and biases of the $l$th module.

\begin{figure}[!tb]
	\centering
	\includegraphics[width=1\linewidth]{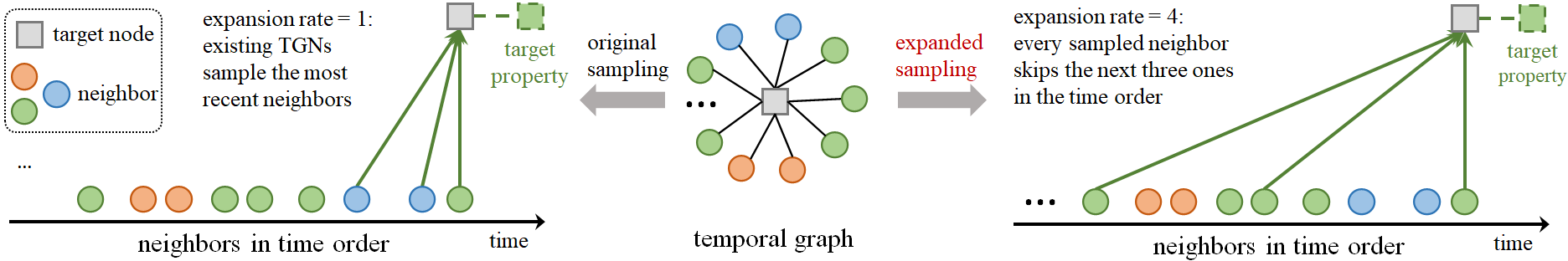}
	\caption{Colors indicate nodes' properties. (\textit{left}) The expanded neighbor sampling with expansion rate as 1 is equivalent the original most recent neighbor sampling. (\textit{right}) The expanded neighbor sampling with expansion rate as 4 aggregates the relevant neighbors' messages to the target node, which is beneficial for predicting the target property.  \label{fig:exp1}}
\end{figure}

\subsection{Expanded Neighbor Sampling} \label{sec:3_2}
Sampling the most recent neighbors as Eq. \eqref{eq:sam} outperforms uniform sampling by utilizing temporal information \cite{rossi2020temporal}. 
However, it has two main limitations.
First, successive
neighbors along the time axis tend to convey redundant features.
For example, a person can interact with another 
multiple times during a short period to discuss a topic that he is interested in.
These repeated and redundant interactions
do not provide more useful
information
and waste
the sampling budget of $S$.
Secondly, 
sampling the most recent neighbors given the budget $S$ limits the temporal scale of each node's
receptive neighborhood, i.e., the neighbors earlier than the $S$th most recent neighbor are inaccessible.

To address the above issues, we propose the expanded neighbor sampling method for TGNs, where every sampled neighbor skips the next neighbors in time order, as visualized in Fig. \ref{fig:exp1}.
Formally, given the expansion ratio of $r$, we have the neighbor sampling indices 
as:
\begin{equation}
\mathcal{S}_i(t) = \{1 + (s - 1) \cdot r\ |\ s \in \{1, \dots, S\},\ 1 + (s - 1) \cdot r \le N(i,t) \},
\end{equation}
where $S$ is the number of sampled neighbors.
Our expanded sampling utilizes the temporal information for neighbor sampling in a flexible way by introducing the expansion rate.
It can reduce the redundancy and skip noisy neighbors by inserting sampling spacing between the sampled neighbors.
In addition, our expanded neighbor sampling expands the temporal scale of the receptive neighborhood, without extra computational load.
The empirical results (see Sec. \ref{sec:4_3}) show that the expanded neighbor sampling performs better than the original neighbor sampling that samples only the most recent neighbors, and simply increasing the sampling budget $S$ for the most recent sampling does not improve the performance by as much as our expanded neighbor sampling approach.

\subsection{Time-aware Neighbor Sampling}
The expanded neighbor sampling fixes a unified expansion rate for all nodes at any time, which may be suboptimal since the predictions on different nodes at different time can require various expansion rates.
For instance, an engineer discusses the project progress with his partners every hour when facing a tight deadline, while he meets his fitness coach for exercise every week on holiday.
In the former case, the expansion rate should be small to aggregate the information from his intensive discussions to predict his working status, while in the latter case, the expansion rate needs to be 
larger to incorporate 
his exercise information to predict his training behaviors.
Enforcing nodes to use the unified expansion rates cannot effectively capture the relevant information and even absorb noise.

\begin{wrapfigure}{r}{7.5cm}\vspace{-5mm}
	\centering
	\includegraphics[width=1\linewidth]{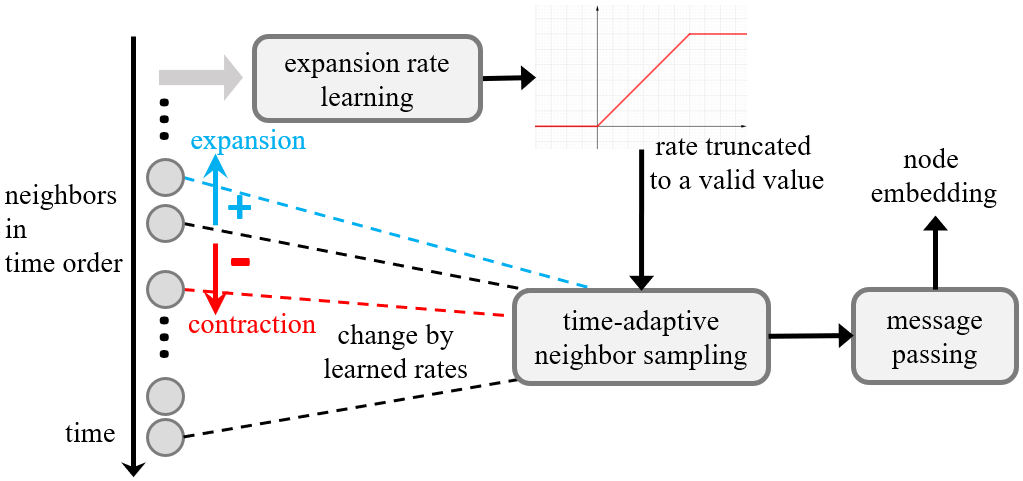}
	\caption{Our TNS learns the expansion rate $r^{(l)}_i(t)$ for any node $i$ at any time $t$ as Eq. \eqref{eq:rat}, and then controls the expansion or contraction of the receptive neighborhood by neighbor sampling as Eq. \eqref{eq:tra}. The black dotted lines indicate the current sampled neighbors. \label{fig:fed}}
	\vspace{-5mm}
\end{wrapfigure}

Hence, beyond a single unified expansion rate, 
we propose \textit{Time-aware Neighbor Sampling} (TNS) to learn the expansion rate for node $i$ at time $t$ and module $l$ through an expansion learning module:
\begin{equation}\label{eq:rat}
r_i^{(l)}(t) = \sigma\Big(\mathsf{AGGR}_r^{(l)}\big(\mathbf{h}_i^{(l-1)}(t), \{\mathbf{m}_{i}^{(l)}(n,t)|n = 1,\dots, S\}\big)\Big),
\end{equation}
where the output dimension of $\mathsf{AGGR}_r^{(l)}$ is $1$.
$\mathsf{AGGR}_r^{(l)}$ follows the implementation of
the original $\mathsf{AGGR}^{(l)}$ introduced in Eq. \eqref{eq:lay}, which is effective on temporal graph learning \cite{rossi2020temporal,xu2020inductive}.
The nonlinear function $\sigma(\cdot)$ truncates the scalar to a valid expansion rate $r_i^{(l)}(t)$, which is defined as 
\begin{equation}\label{eq:tru}
\sigma(x) = 	
\begin{cases}
\min\big(\max(1, x), (N(i,t) - 1) / (S - 1)\big) & \mathrm{if}\ N(i,t) \ge S \\
1 & \mathrm{otherwise}.
\end{cases}
\end{equation}
Then, the expanded neighbor sampling indices for node $i$ at time $t$ and module $l$ are defined as:
\begin{equation}\label{eq:tra}
\mathcal{S}^{(l)}_i(t) = \big\{1 + (s - 1) \cdot r_i^{(l)}(t)\ |\ s = 1, \dots, S,\ 1 + (s - 1) \cdot r_i^{(l)}(t) \le N(i,t)\big\}.
\end{equation}
The indices in $\mathcal{S}_i^{(l)}(t)$ may not be integers since $r_i^{(l)}(t)$ is likely to not be an integer.
To make neighbors of any index in time order accessible, we compare the neighbor sampling to 
the process of image rendering  \cite{kirillov2020pointrend},
where we view neighbors as image pixels.
Concretely, we perform 
the interpolations for the neighbors located on $n\in \mathcal{S}_i^{(l)}(t)$ as following:
\begin{equation}\label{eq:int}
\mathbf{m}_{i}^{(l)}(n,t) = \sum_{o=1}^{N(i,t)}\max\big(0, 1 - |n - o|)\cdot \mathbf{m}_{i}^{(l)}(o, t), \forall n \in \mathcal{S}_i^{(l)}(t),
\end{equation}
which is fast to compute as $\max(0, 1 - |n - o|)$ is non-zero for at most two
$i \in\{ 1,\dots, N(i,t)\}$.
The feed-forward of the temporal graph aggregation with our time-aware neighbor sampling is:
\begin{equation}
\mathbf{h}_i^{(l)}(t) = \mathsf{AGGR}^{(l)}\Big(\mathbf{h}_i^{(l-1)}(t), \big\{\mathbf{m}_{i}^{(l)}(n, t)\ |\ n \in  \mathcal{S}^{(l)}_i(t)\big\}\Big),
\end{equation}
which is visualized in Fig. \ref{fig:fed}.
The expansion rate $r_i^{(l)}(t)$ in Eq. \eqref{eq:rat} is learned to control the temporal scale of the receptive neighborhood of node $i$ at time $t$ and module $l$.
A increase in $r_i^{(l)}(t)$'s 
implies the expansion of the receptive neighborhood, while a decrease indicates the contraction.


Next, we analyze how TNS is implicitly and automatically guided to sample the neighbors that are most beneficial for prediction.
Denote the target loss as $\mathcal{L}$, e.g., cross-entropy for node classification \cite{rossi2020temporal}.
On backward propagation, 
the partial derivatives of the neighbor index $n\in\mathcal{S}_i^{(l)}(t)$ can be obtained according to the equations:
\begin{equation}
\frac{\partial \mathcal{L}}{\partial n} = \sum_{o=1}^{N(i,t)} \Big(\frac{\partial \mathcal{L}}{\partial \mathbf{m}_{i}^{(l)}(n,t)}\Big)^T \mathbf{m}_{i}^{(l)}(o,t) \cdot \mathbbm{1}(|n - o| < 1) \cdot \mathrm{sgn}(o - n),\ \forall n \in \mathcal{S}_i^{(l)}(t)
\end{equation}
where the function $\mathrm{sgn}(o - n)$ returns $1$ if $o > n$, and $-1$ otherwise.
The function $\mathbbm{1}(|n - o| < 1)$ returns $1$ if $|n - o| < 1$, and $0$ otherwise.
The signum of the gradient $\frac{\partial \mathcal{L}}{\partial n}$ controls the expansion (-1) or contraction (1) on the neighbor index $n \in \mathcal{S}_i^{(l)}(t)$ (when $n$ is a decimal and $1 \le n \le N(i,t)$):
\begin{align}\label{eq:sgn}
\mathrm{sgn}\Big(\frac{\partial \mathcal{L}}{\partial n}\Big) = &\ \mathrm{sgn}\bigg(\sum_{o=1}^{N(i, t)} \Big(\frac{\partial \mathcal{L}}{\partial \mathbf{m}_{i}^{(l)}(n,t)}\Big)^T \mathbf{m}_{i}^{(l)}(o,t) \cdot \mathbbm{1}(|n - o| < 1) \cdot \mathrm{sgn}(o - n)\bigg) \nonumber \\
=&\ \mathrm{sgn}\bigg(\Big(\frac{\partial \mathcal{L}}{\partial \mathbf{m}_{i}^{(l)}(n,t)}\Big)^T \mathbf{m}_{i}^{(l)}(\lfloor n\rfloor + 1,t) - \Big(\frac{\partial \mathcal{L}}{\partial \mathbf{m}_{i}^{(l)}(n, t)}\Big)^T \mathbf{m}_{i}^{(l)}(\lfloor n\rfloor, t)\bigg) \nonumber\\
=&\ \mathbbm{1}\bigg(\Big(\frac{\partial \mathcal{L}}{\partial \mathbf{m}_{i}^{(l)}(n,t)}\Big)^T \mathbf{m}_{i}^{(l)}(\lfloor n \rfloor + 1, t) > \Big(\frac{\partial \mathcal{L}}{\partial \mathbf{m}_{i}^{(l)}(n,t)}\Big)^T \mathbf{m}_{i}^{(l)}(\lfloor n \rfloor,t)\bigg),
\end{align}
where $\lfloor n\rfloor$ is the maximum integer smaller than $n\in \mathcal{S}_i^{(l)}(t)$.
Eq. \eqref{eq:sgn} shows that the expansion or contraction on $n \in \mathcal{S}_i^{(l)}(t)$ is determined by the comparison between the projections of $\mathbf{m}_{i}^{(l)}(\lfloor n \rfloor, t)$ and $\mathbf{m}^{(l)}_{i}(\lfloor n \rfloor + 1,t)$ on $\frac{\partial \mathcal{L}}{\partial \mathbf{m}_{i}^{(l)}(n,t)}$.
$\frac{\partial \mathcal{L}}{\partial \mathbf{m}_{i}^{(l)}(n,t)}$ represents the direction of message $\mathbf{m}_{i}^{(l)}(n,t)$ that increases $\mathcal{L}$.
If the projection on $\frac{\partial \mathcal{L}}{\partial \mathbf{m}_{i}^{(l)}(n,t)}$ of $\mathbf{m}_{i}^{(l)}(\lfloor n \rfloor, t)$ is smaller than that of $\mathbf{m}_{i}^{(l)}(\lfloor n \rfloor + 1, t)$, the $\lfloor n \rfloor + 1$th neighbor gives more strength on loss increasing than the $\lfloor n \rfloor$th neighbor, i.e., the neighbor $\lfloor n \rfloor$ is more beneficial to the prediction of the target node $i$.
In this case, $\mathrm{sgn}(\frac{\partial \mathcal{L}}{\partial n}) = 1$ contracts $n$ towards the more beneficial $\lfloor n \rfloor$th neighbor.
In the other case, $\mathrm{sgn}(\frac{\partial \mathcal{L}}{\partial n}) = -1$ expands $n$ towards the more beneficial neighbor $\lfloor n \rfloor + 1$th neighbor.
Both cases meet our expectation.
Overall, the gradient $\frac{\partial \mathcal{L}}{\partial n}$ guides $n$ to move in the direction that is most beneficial for prediction.

\subsection{Analysis and Discussion}\label{sec:34}
Initialization of the expansion learning modules $\mathsf{AGGR}_r^{(l)}$ in Eq. \eqref{eq:rat} is crucial to give reasonable expansion rates at early training stages and reduce optimization difficulty.
We initialize the output-layer weight and bias of $\mathsf{AGGR}_r^{(l)}$, e.g., $\mathbf{W}_2^{(l)}$ and $\mathbf{b}_2^{(l)}$ in Eq. \eqref{eq:lay}, as small values close to zero following $\mathcal{N}(0, \sigma^2), \sigma \ll 1$, and $1$ respectively.
Here, $\mathcal{N}$ is the normal distribution \cite{ahsanullah2014normal}.
The advantages of our initialization method are three-fold.
First, the initialized expansion rates are close to 1, i.e., TNS will start from the original neighbor sampling and gradually learn the appropriate expansion rates during training.
Second, by setting the weight elements to small values, i.e., $\sigma \ll 1$, the expansion rates are insensitive to the input representations at the early training stage, which lack effective semantic information.
Third, our initialization generates the expansion rates close to 1 before the truncation function $\sigma(\cdot)$ in Eq. \eqref{eq:rat}, which prevents the optimizer from taking large efforts to push the ill-conditioned expansion rates to the activation range $[1, (N(i,t) - 1) / (S - 1)]$ of $\sigma(\cdot)$, as shown in Eq. \eqref{eq:tru}.
Otherwise, if the expansion rates before $\sigma(\cdot)$ are far away from the activation range $[1, (N(i,t) - 1) / (S - 1)]$ of $\sigma(\cdot)$, $\mathsf{AGGR}_r^{(l)}$ receives
only zero gradients.

Besides the reasonable initialization, we set the learning rate of the expansion learning modules $\mathsf{AGGR}_r^{(l)}$ as $\alpha$ times of the learning rate for the original TGNs' existing modules.
In our implementation, we set $\alpha < 1$ to make the expansion learning modules updated slower than others.
This makes the neighbor indices updated stably from the reasonably initialized expansion rates in the early training stages and the appropriate expansion rates in the later stages.

Our expanded neighbor sampling and TNS can be incorporated into advanced TGN models and are applicable to many tasks on the temporal graph data, such as node classification and edge prediction \cite{rossi2020temporal}, \cite{xu2020inductive}.
The expansion rates of the former
, as hyper-parameters, can be found by 
the grid search on the validation data, while our TNS learns the adaptive expansion rates automatically and is trained in an end-to-end manner.

In terms of the time complexity, our expanded neighbor sampling does not induce any extra computation load and thus does not increase the original time complexity.
Our TNS introduces the expansion learning module $\mathsf{AGGR}_r^{(l)}$, which induces less computation load than the corresponding temporal graph aggregation module $\mathsf{AGGR}^{(l)}$, since the output dimension of $\mathsf{AGGR}_r^{(l)}$ is fixed as 1.
Therefore, TNS does not increase the time complexity as well.
Taking the mean aggregation in Eq. \eqref{eq:lay} as an example, we have the time complexity of the original TGNs' $\mathsf{AGGR}^{(l)}$ as $\mathcal{O}(d_m \cdot (S + d_h) + d_o \cdot (d_i + d_h))$, where $d_m$, $d_h$, $d_i$, $d_o$ are the dimension of messages $\mathbf{m}_{n}^{(l)}(i, t)$, and representations $\hat{\mathbf{h}}_i^{(l)}(t)$, $\mathbf{h}_i^{(l - 1)}(t)$, $\mathbf{h}_i^{(l)}(t)$ respectively.
With the output dimension as 1, the expansion learning module $\mathsf{AGGR}_r^{(l)}$ has the complexity $\mathcal{O}(d_m \cdot (S + d_h) + d_i + d_h)$.
The complexity of $\mathsf{AGGR}_r^{(l)}$ and $\mathsf{AGGR}^{(l)}$ together is $\mathcal{O}(d_m \cdot (S + d_h) + d_o \cdot (d_i + d_h))$, which is as same as the original TGNs.


\section{Experiments}
In this section, we present the performance of TGN models implemented with our neighbor sampling methods.
We compare our proposed method against a variety of strong baselines (adapted for temporal settings when possible) on the task of edge prediction and node classification.
Our experimental settings closely follow those of the previous work \cite{rossi2020temporal,xu2020inductive,wang2020apan} to ensure a fair comparison.
As for the evaluation metrics, we follow \cite{xu2020inductive,wang2020apan} to use the average precision (AP) and accuracy in edge prediction and employ the area under the ROC accuracy curve (AUC) for node classification \cite{kumar2019predicting}.

We use three standard temporal graph datasets: MOOC, Reddit, and Wikipedia for evaluation.
The MOOC dataset consists of actions, e.g., viewing a video, submitting an answer, etc. done by students on a MOOC online course \cite{kumar2019predicting}.
The Reddit dataset consists of one month of posts made by users on subreddits \cite{baumgartner2020pushshift}.
The user posts have textual features that are transformed into a 172-dimensional vector representing its linguistic inquiry and word count (LIWC) categories.
The Wikipedia dataset includes one month of edits on Wikipedia pages \cite{wiki}.
The statistics of these datasets are shown in Table \ref{tab:data}.
We do the
chronological train-validation-test split with a 
ratio of 70\%-15\%-15\% following \cite{rossi2020temporal,xu2020inductive}.

For the hyper-parameters of baseline methods, e.g., the number of sampled neighbors, the number of hidden units, the optimizer, the number of modules, the number of sampled neighbors, and the learning rate, we set them as suggested by their authors.
For the hyper-parameters of our TNS method, we set the gradient factor of the expansion learning module as $\alpha = 0.1$, and the standard deviation $\sigma = 10^{-5}$ for the initialization of the expansion learning module by default.

\begin{table}[tb!]
	\centering
	\caption{Statistics of the datasets used in our experiments. Edges refer to temporal edges.}
	\label{tab:data}
	
	\begin{adjustbox}{width=\linewidth}
		
	\begin{tabular}{@{}l| c c c c c c@{}}
		\toprule
		\textbf{Dataset}
		& $\#$\textbf{Nodes}
		& $\#$\textbf{Edges}
		& \textbf{$\#$Nodes in val./test.}
		& \textbf{$\#$Edges in val./test.}
		& \textbf{$\#$Nodes with dynamic labels}
		& \textbf{Nodes' label type} \\ 
		\midrule
		\midrule
		\texttt{MOOC} & 7,144 & 411,749 & 2,599/2,412 & 61,762/61,763 & 4,066 & course dropout\\
		\texttt{Reddit} & 10,984 & 672,447 &  9,839/9,615 & 100,867/100,867 & 366 & posting ban\\
		\texttt{Wikipedia} & 9,227 & 157,474 &  3,256/3,564 & 23,621/23,621 & 217 & editing ban\\
		\bottomrule
	\end{tabular}

	\end{adjustbox}

\end{table}

\begin{table}[tb!]
	\centering
	\caption{Test accuracy and average precision (AP) of transductive edge prediction.
		We conduct 100 trials with random weight initialization. Mean (\%) and standard deviations are reported. The best results in each column are highlighted in \textbf{bold} font.}
	\label{tab:trans}
	
	\begin{adjustbox}{width=\linewidth}
	
	\begin{tabular}{@{}l|cc|cc|cc@{}}
		\toprule 
		\multirow{2}{*}{\textbf{Method}} & \multicolumn{2}{c|}{\texttt{MOOC}} & \multicolumn{2}{c|}{\texttt{Reddit}} & \multicolumn{2}{c}{\texttt{Wikipedia}} \\ 
		
		& Accuracy & AP &  Accuracy & AP &  Accuracy & AP \\
		\midrule
		\midrule
		CTDNE \cite{nguyen2018continuous} & 65.34 $\pm$ 0.7 & 74.29 $\pm$ 0.6 & 73.76 $\pm$ 0.5 & 91.41 $\pm$ 0.3 & 79.42 $\pm$ 0.4 & 92.17 $\pm$ 0.5 \\ 
		JODIE \cite{kumar2019predicting} & 76.45 $\pm$ 0.6 & 83.87 $\pm$ 0.4 & 90.91 $\pm$ 0.3 & 97.11 $\pm$ 0.3 & 87.04 $\pm$ 0.4 & 94.62 $\pm$ 0.5 \\ 
		DyRep \cite{trivedi2019dyrep} & 73.36 $\pm$ 0.4 & 81.75 $\pm$ 0.3 & 92.11 $\pm$ 0.2 & 97.98 $\pm$ 0.1 & 87.77 $\pm$ 0.2 & 94.59 $\pm$ 0.2 \\
		TGAT \cite{xu2020inductive} & 75.20 $\pm$ 0.5 & 82.66 $\pm$ 0.4 & 92.92 $\pm$ 0.3 & 98.12 $\pm$ 0.2 & 88.14 $\pm$ 0.2 & 95.34 $\pm$ 0.1 \\ 
		TGN \cite{rossi2020temporal} & 81.38 $\pm$ 0.6 & 89.79 $\pm$ 0.5 & 92.56 $\pm$ 0.2 & 98.70 $\pm$ 0.1 & 89.51 $\pm$ 0.4 & 98.46 $\pm$ 0.1 \\ 
		\midrule
		TGAT + TNS (Ours) & 80.76 $\pm$ 0.5 & 89.20 $\pm$ 0.3 & 93.95 $\pm$ 0.3 & 98.68 $\pm$ 0.1 & 89.59 $\pm$ 0.2 & 96.73 $\pm$ 0.1\\
		TGN + TNS (Ours) & \textbf{84.42 $\pm$ 0.3} & \textbf{92.54 $\pm$ 0.3} & \textbf{94.04 $\pm$ 0.2} & \textbf{98.93 $\pm$ 0.1} & \textbf{91.61 $\pm$ 0.2} & \textbf{98.93 $\pm$ 0.1}\\
		\bottomrule
	\end{tabular}

	\end{adjustbox}

\end{table}

\begin{table*}[tb!]
	\centering
	\caption{Test accuracy and average precision (AP) of inductive edge prediction.
		We conduct 100 trials with random weight initialization. Mean (\%) and standard deviations are reported. The best results in each column are highlighted in \textbf{bold} font.}
	\label{tab:ind}
	
	\begin{adjustbox}{width=\linewidth}
		
		\begin{tabular}{@{}l|cc|cc|cc@{}}
			\toprule 
			\multirow{2}{*}{\textbf{Method}} & \multicolumn{2}{c|}{\texttt{MOOC}} &	        \multicolumn{2}{c|}{\texttt{Reddit}} & \multicolumn{2}{c}{\texttt{Wikipedia}} \\ 
			
			& Accuracy & AP &  Accuracy & AP &  Accuracy & AP\\
			\midrule
			\midrule
			JODIE \cite{kumar2019predicting} & 75.79 $\pm$ 0.5 & 83.44 $\pm$ 0.6 & 88.34 $\pm$ 0.9 & 94.36 $\pm$ 1.1 & 84.32 $\pm$ 0.4 & 93.11 $\pm$ 0.4 \\ 
			DyRep \cite{trivedi2019dyrep} & 72.92 $\pm$ 0.4 & 80.36 $\pm$ 0.4 & 89.60 $\pm$ 0.2 & 95.68 $\pm$ 0.2 & 83.46 $\pm$ 0.3 & 92.05 $\pm$ 0.3 \\ 
			TGAT \cite{xu2020inductive} & 74.02 $\pm$ 0.3 & 80.84 $\pm$ 0.5 & 90.73 $\pm$ 0.2 & 96.62 $\pm$ 0.3 & 85.35 $\pm$ 0.2 & 93.99 $\pm$ 0.3 \\
			TGN \cite{rossi2020temporal} & 80.73 $\pm$ 0.2 & 89.21 $\pm$ 0.3 & 91.62 $\pm$ 0.1 & 97.55 $\pm$ 0.1 & 88.60 $\pm$ 0.2 & 97.81 $\pm$ 0.1 \\ 
			\midrule
			TGAT + TNS (Ours) & 80.45 $\pm$ 0.3  & 87.82 $\pm$ 0.4 & 91.89 $\pm$ 0.2 & 97.09 $\pm$ 0.2 & 87.27 $\pm$ 0.2 & 95.84 $\pm$ 0.3\\
			TGN + TNS (Ours) & \textbf{83.81 $\pm$ 0.2} & \textbf{91.36 $\pm$ 0.3} & \textbf{92.71 $\pm$ 0.1} & \textbf{98.02 $\pm$ 0.1} & \textbf{91.14 $\pm$ 0.2} & \textbf{98.42 $\pm$ 0.1}\\
			\bottomrule
		\end{tabular}
		
	\end{adjustbox}
	
\end{table*}

\subsection{Edge Prediction}\label{sec:edge}
Following \cite{xu2020inductive,rossi2020temporal}, we conduct the experiments for edge prediction under both transductive and inductive settings for a comprehensive evaluation.
In the transductive task, we predict future edges of the nodes observed during training, whereas in the inductive one we predict future edges of nodes never observed during training.

In the transductive edge prediction, we take the state-of-the-art approaches for the representation learning on temporal graphs: CTDNE \cite{nguyen2018continuous}, JODIE \cite{kumar2019predicting}, DyRep \cite{trivedi2019dyrep}, TGAT \cite{xu2020inductive}, TGN \cite{rossi2020temporal} as the baselines for comparison \cite{xu2020inductive}.
We conduct the experiments for 100 trials with random weight initialization.
We implement TSN with the TGN models TGAT and TGN.
Table \ref{tab:trans} reports the results.
Our TNS method improves the test accuracy of TGAT by 7.4\% on MOOC, 1.1\% on Reddit, 1.6\% on Wikipedia, and improves the test accuracy of TGN by 3.7\% on MOOC, 1.6\% on Reddit, 2.3\% on Wikipedia.
In terms of AP, TNS achieves similar improvements.
As a result, our TNS achieves substantial improvements for TGAT and TGN.

In the inductive setting, we keep the baselines which support
inductive learning for comparison.
We conduct the experiments for 100 trials with random weight initialization.
The results are reported in Table \ref{tab:ind}.
We implement our TNS method with TGAT and TGN to study whether TNS can improve the performance of TGNs under the inductive setting.
We observe that TNS improves the test accuracy of TGAT by 8.7\% on MOOC, 1.3\% on Reddit, 2.2\% on Wikipedia, and TGN by 3.8\% on MOOC, 1.2\% on Reddit, 2.9\% on Wikipedia.
As a result, our TNS method enhances TGAT and TGN to outperform the baseline methods in the inductive task.

Given the TGN models TGAT and TGN, our TNS achieves consistent and substantial improvements on all datasets, thanks to the adaptive receptive neighborhood offered by our TNS method.
TNS effectively learns the appropriate expansion rate for different nodes at different timestamps in an end-to-end style.
Overall, the results above indicate that our approach is effective in improving the effectiveness of the popular TGN models on both transductive and inductive settings.

\subsection{Node Classification}

\begin{wraptable}{r}{8.5cm}\vspace{-4mm}
	\centering
	\caption{ROC AUC (\%) on the test set for temporal node classification. We conduct 100 trials with random weight initialization. Mean (\%) and standard deviations are reported.}
	\label{tab:node}
	
	\begin{adjustbox}{width=\linewidth}
			
	\begin{tabular}{@{}lccc@{}}
		\toprule
		\textbf{Method} & \texttt{MOOC} & \texttt{Reddit} &\texttt{Wikipedia} \\ 
		\midrule
		\midrule
		CTDNE \cite{nguyen2018continuous} & 67.54 $\pm$ 0.7 & 59.43 $\pm$ 0.6 & 75.89 $\pm$ 0.5 \\ 
		JODIE \cite{kumar2019predicting} & 76.31 $\pm$ 1.6 & 61.83 $\pm$ 2.7 & 84.84 $\pm$ 1.2 \\ 
		DyRep \cite{trivedi2019dyrep} & 75.32 $\pm$ 1.3 & 62.91 $\pm$ 2.4 & 84.59 $\pm$ 2.2\\ 
		TGAT \cite{xu2020inductive} & 74.25 $\pm$ 0.9 & 65.56 $\pm$ 0.7 & 83.69 $\pm$ 0.7 \\ 
		TGN \cite{rossi2020temporal} & 77.73 $\pm$ 0.7 & 67.06 $\pm$ 0.9 & 87.81 $\pm$ 0.3 \\ 
		\midrule
		TGAT + TNS (Ours) & 75.52 $\pm$ 1.0 & 66.07 $\pm$ 1.1 & 85.32 $\pm$ 0.7\\
		TGN + TNS (Ours) & \textbf{79.12 $\pm$ 0.6} & \textbf{68.69 $\pm$ 0.9} & \textbf{89.04 $\pm$ 0.4}\\
		\bottomrule
	\end{tabular}

	\end{adjustbox}
\vspace{-5mm}
\end{wraptable}

The task of node classification on temporal graphs is to predict the time-varying labels of nodes.
The meaning of node labels varies in different datasets.
For example, in Reddit, node labels correspond to the banning of posting on the Reddit platform \cite{baumgartner2020pushshift}.

Table \ref{tab:node} reports the test ROC AUC.
We observe that TNS improves ROC AUC of TGAT by 1.7\% on MOOC, 0.8\% on Reddit, 1.9\% on Wikipedia, and TGN by 1.8\% on MOOC, 2.4\% on Reddit, 1.4\% on Wikipedia.
As a result, our TNS method enhances TGAT and TGN to outperform the baseline methods in the temporal node classification.
This validates the importance of appropriate receptive neighborhoods for guiding temporal graph learning, which adaptively aggregates the relevant information to the target node.

\begin{figure}[!tb]
	\centering
	\begin{subfigure}[t]{0.45\textwidth}
		\includegraphics[width=\textwidth]{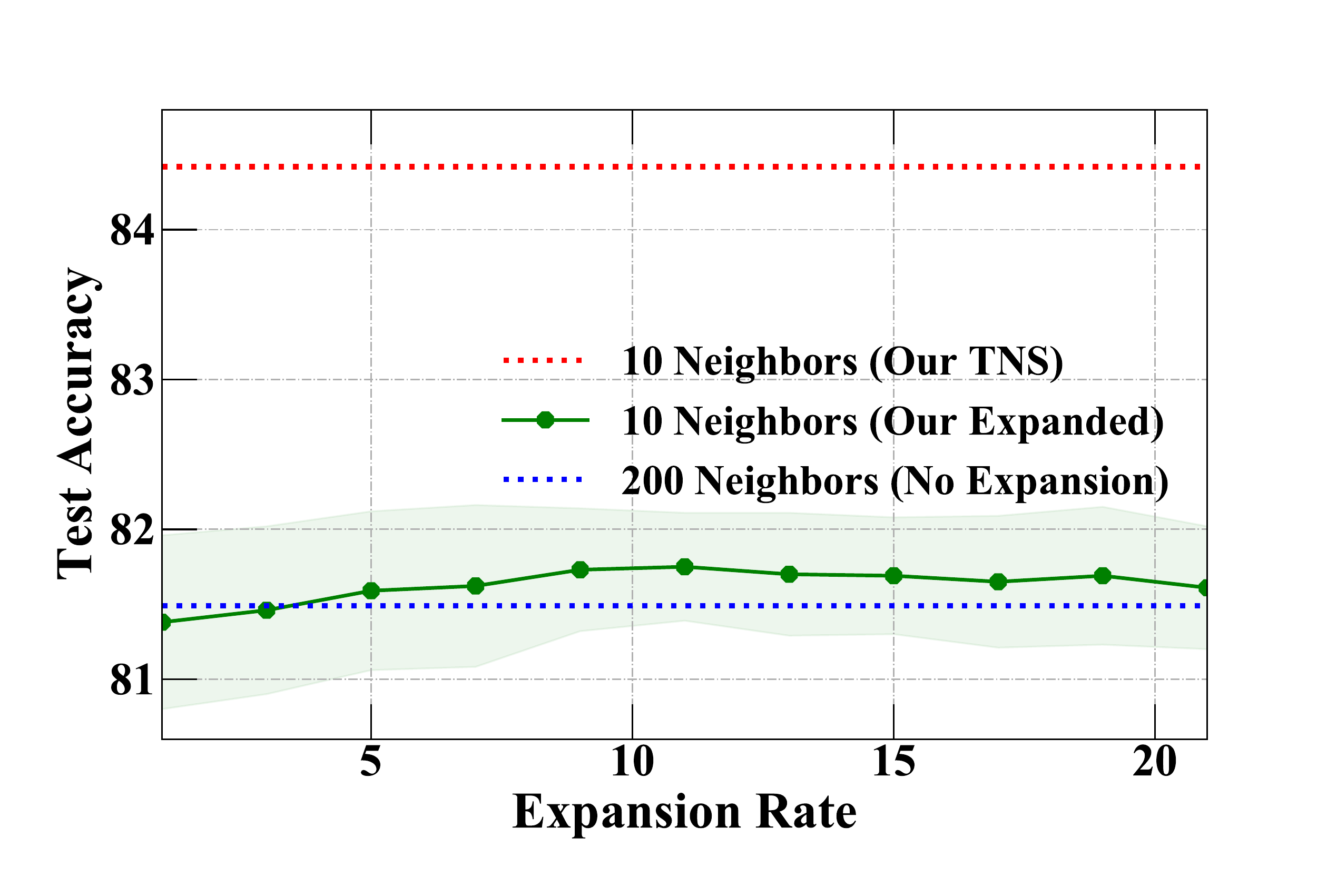}
	\end{subfigure}
	\begin{subfigure}[t]{0.45\textwidth}
		\includegraphics[width=\textwidth]{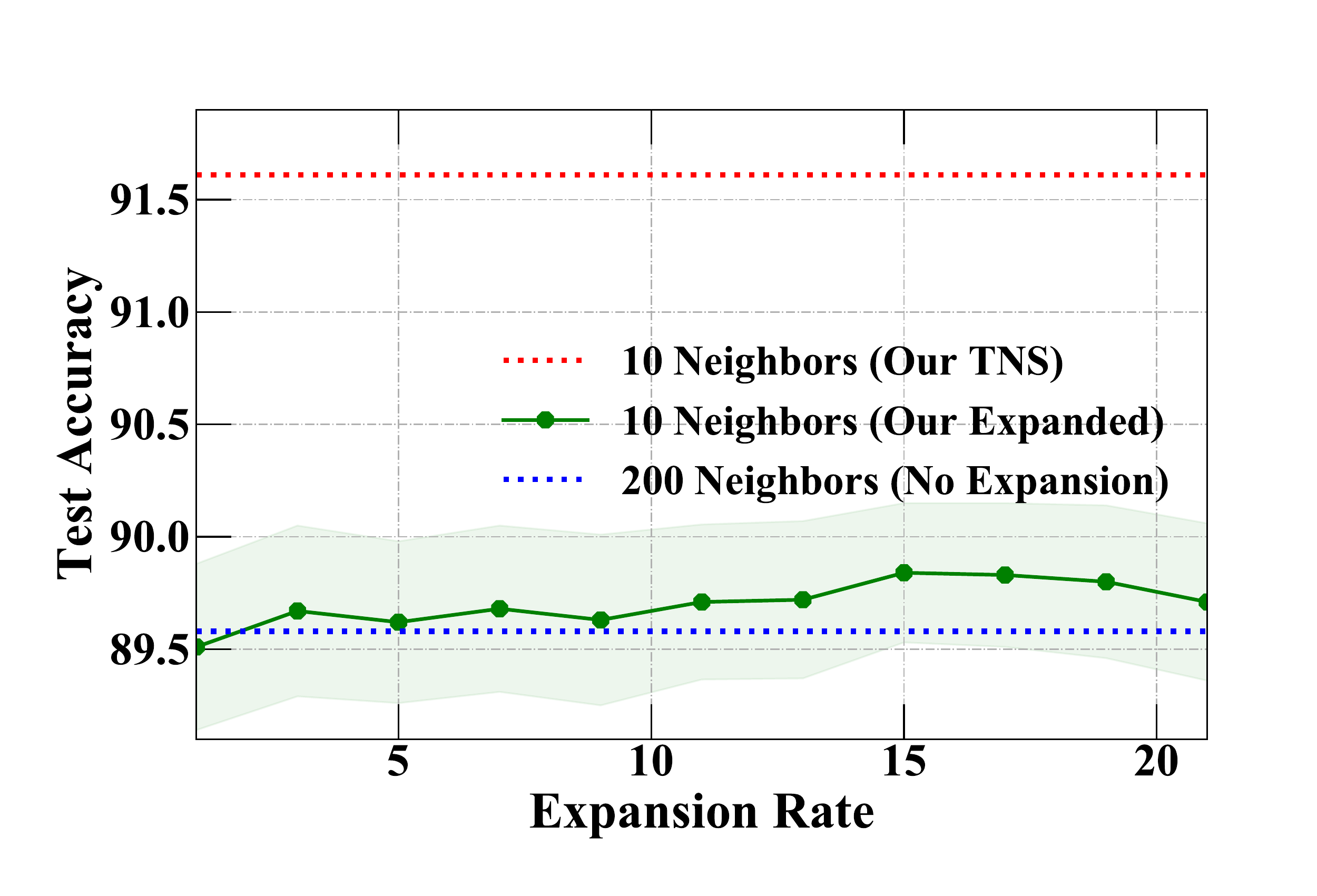}
	\end{subfigure}	\hfill
	\caption{(\textit{left, right}) The test accuracy of TGN \cite{rossi2020temporal} on edge prediction versus expansion rates on the MOOC and Wikipedia datasets respectively. The performance of TGN changes slightly with the expansion rate increasing from 1 to 21, while our TNS exhibits significant improvements given the same number sampled neighbors. \label{fig:exp2}}
	\vspace{-5mm}
\end{figure}


\subsection{Expanded Neighbor Sampling}\label{sec:4_3}
We compare the performance of our expanded neighbor sampling introduced in Sec. \ref{sec:3_2} with the original most recent sampling and our TNS method.
We implement TGN \cite{rossi2020temporal} with them and follow the experimental settings of the transductive edge prediction introduced in Sec. \ref{sec:edge}.
Note that the default number of the sampled neighbors suggested by \cite{rossi2020temporal} is 10.
Fig. \ref{fig:exp2} presents the test accuracy on the MOOC and Wikipedia datasets.
The expanded sampling with the expansion rate equal to 1 degrades to the original most recent sampling.
The expanded neighbor sampling performs better than the unexpanded one, since it reduces the information redundancy and some nodes demand larger receptive temporal neighborhoods for prediction.
Note that with extremely large expansion rates, the performance can degrade because fewer neighbors are sampled.
Interestingly, compared with the unexpanded sampling with the number of neighbors increasing from 10 to 200 with higher computation load and larger receptive temporal neighborhood, our expanded sampling still performs better. 
This validates that the advantages of our expanded sampling on reducing the information redundancy and denoising the input messages.

Taking a closer look, we find that the improvements achieved by our TNS are 
much higher than that given by our expanded sampling of different rates. 
Considering that the overlap between the set of sampled neighbors with different expansion rates are small, this result implies that different nodes can demand various expansion rates at different time, and the adaptive temporal neighborhood can lead to 
significant improvements of TGNs.


\begin{table}[!tb]
	\centering
	\caption{Test accuracy of inductive edge prediction with different numbers of sampled neighbors $S$.
    Mean (\%) and standard deviations are reported.}
	\label{tab:num}
	
	\begin{adjustbox}{width=\linewidth}
	\begin{tabular}{@{}l|ccc|ccc@{}}
		\toprule 
		\multirow{2}{*}{\textbf{Method}} &
		\multicolumn{3}{c|}{\texttt{MOOC}} & \multicolumn{3}{c}{\texttt{Wikipedia}} \\
		& $S = 10$ & $S = 20$ &  $S = 50$
		& $S = 10$ &  $S = 20$ & $S = 50$\\
		\midrule
		\midrule
		TGAT \cite{xu2020inductive} & 73.58 $\pm$ 0.2 & 74.02 $\pm$ 0.3 & 74.10 $\pm$ 0.4 & 84.52 $\pm$ 0.2 & 85.35 $\pm$ 0.2 & 85.52 $\pm$ 0.3 \\ 
		TGAT + TNS (Ours) & \textbf{80.17 $\pm$ 0.2} & \textbf{80.45 $\pm$ 0.3} & \textbf{80.51 $\pm$ 0.3} &  \textbf{86.58 $\pm$ 0.2} & \textbf{87.27 $\pm$ 0.2} & \textbf{87.39 $\pm$ 0.2} \\ 
		\midrule
		TGN \cite{rossi2020temporal} & 80.73 $\pm$ 0.2 & 80.86 $\pm$ 0.3 & 80.88 $\pm$ 0.3 & 88.60 $\pm$ 0.2 & 88.68 $\pm$ 0.2 & 88.71 $\pm$ 0.3 \\ 
		TGN + TNS (Ours) & \textbf{83.81 $\pm$ 0.2} & \textbf{83.92 $\pm$ 0.2} & \textbf{83.94 $\pm$ 0.3} &  \textbf{91.14 $\pm$ 0.2} & \textbf{91.18 $\pm$ 0.2} & \textbf{91.20 $\pm$ 0.2}\\
		\bottomrule
	\end{tabular}

	\end{adjustbox}

\end{table}

\begin{wrapfigure}{r}{5cm}\vspace{-5mm}
	\centering
	\includegraphics[width=1\linewidth]{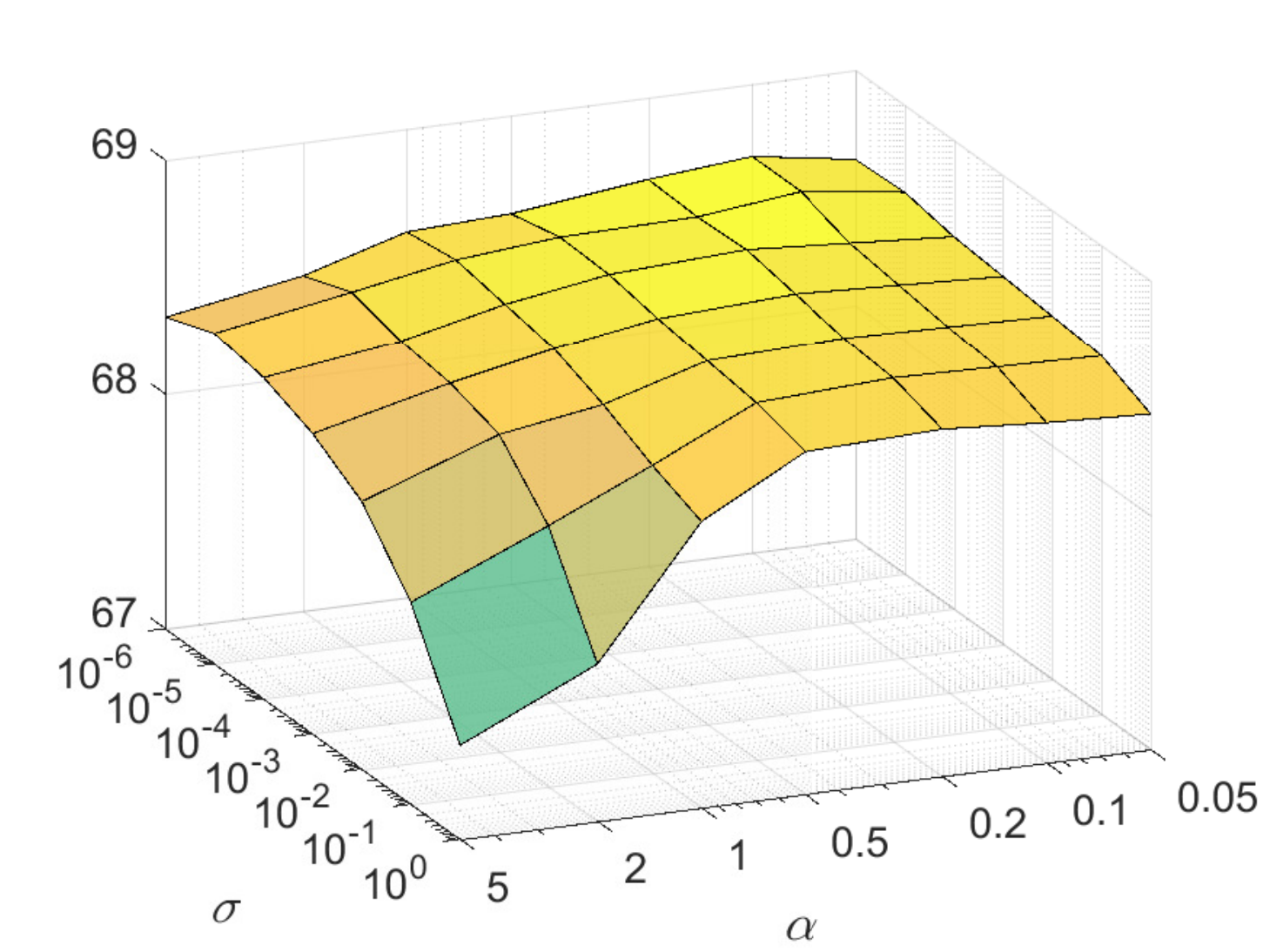}
	\caption{The ROC AUC (\% in z-axis) of TGN with TNS on node classification of 
		Reddit with different hyper-parameters $\alpha$ and $\sigma$.\label{fig:hyperparameter}}
	\vspace{-5mm}
\end{wrapfigure}

\subsection{Ablation Study}
We vary the number of sampled neighbors to observe how the performance of TGNs with or without our TNS changes
with different number of sampled neighbors.
We report the results on node classification in Table \ref{tab:num}.
As the number of sampled neighbors increases 
from $S = 10$ to $S = 50$, both the performance of TGAT and TGN improve, since more neighbors enrich the input information.
For the TGAT and TGN with 
different number of sampled neighbors, our TNS consistently achieves the performance enhancements.
Taking a closer look, we find that the improvements given by TNS are 
much larger than that given by increasing the number of sampled neighbors, since more sampled neighbors without our time-aware sampling cannot reduces noise and redundancy.
In contrast, our TNS method utilizes the temporal information for neighbor sampling in a learnable way, which offers the adaptive receptive neighborhood to every node at any time and can reduce noise and redundancy.

Finally, we investigate the sensitivity of TNS to the hyper-parameters: $\alpha$ and $\sigma$ to control the learning rate and the initialization respectively for the expansion learning modules.
The result is visualized in Fig. \ref{fig:hyperparameter}.
We alter $\alpha$ among $\{0.05, 0.1, 0.5. 1, 2, 5\}$ and $\sigma$ among $\{10^{-6}, 10^{-6}, 10^{-5}, 10^{-4}, 10^{-3}, 10^{-2}, 0.1, 1\}$.
The performance of TGN with TNS is relatively smooth when parameters are within certain ranges.
However, extremely large values of $\alpha$ and small $\sigma$ result in poor performances.
Too large learning rates with large $\alpha$ make the learning of the expansion rates unstable, while too large $\sigma$ cannot offer reasonable initialization to the expansion rates, which should be avoided in practice.
Moreover, only a poorly set hyper-parameter does not lead to significant performance degradation, which demonstrates that our time-aware neighbor sampling method is able to mine the most valuable information from the temporal neighborhood in the end-to-end training.
More experimental results about our methods can be found in Appendix.

\section{Conclusion}\label{sec:con}
In this paper, we have developed a new neighbor sampling method named TNS.
TNS learns how to sample neighbors through an expansion learning module, which is automatically and implicitly guided to move the neighbor indices (in time order) in the direction that is most beneficial for prediction.
TNS offers adaptive receptive neighborhoods to every node at any time to improve the effectiveness of TGNs without increasing their time complexity.
TNS needs no extra supervision and can be trained in an end-to-end manner.
We demonstrate the effectiveness of TNS on datasets comprising 
documents, music, courses, and online discussion posts.
TNS enhances the popular TGN models significantly and enables them to outperform the baseline methods.
A limitation of TNS is that TNS learns how to sample neighbors based on our expanded sampling, which may not be the best option on utilizing the temporal information, so a possible direction of the future work is to explore more flexible methods for neighbor sampling on temporal graphs.

\bibliographystyle{plainnat}
\bibliography{main}


\newpage
\appendix
\section{Appendix}
\subsection{Additional Experiments}
\begin{figure}[!tb]
	\centering
	\begin{subfigure}[t]{0.33\textwidth}
		\includegraphics[width=\textwidth]{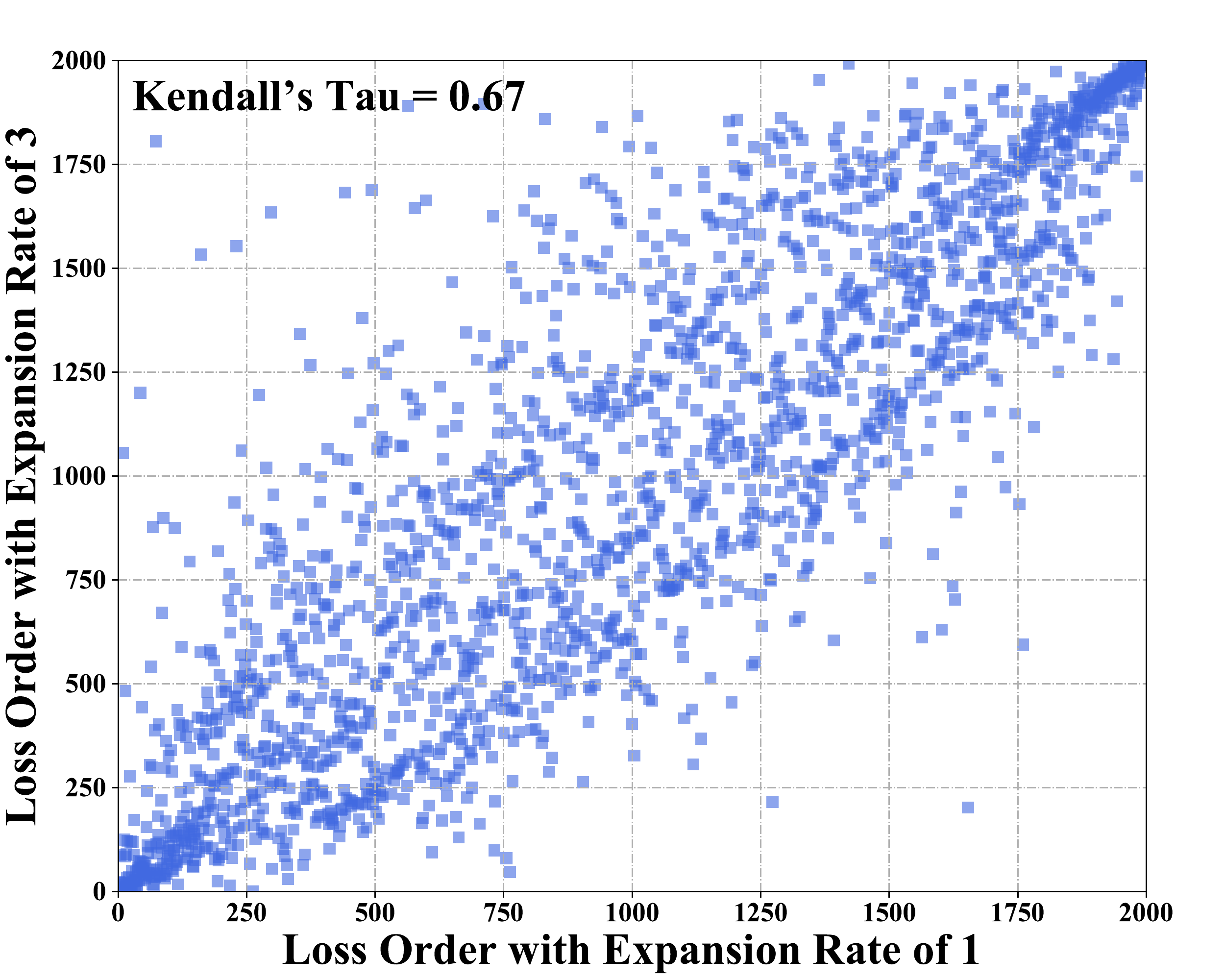}
	\end{subfigure}\hfill
	\begin{subfigure}[t]{0.33\textwidth}
		\includegraphics[width=\textwidth]{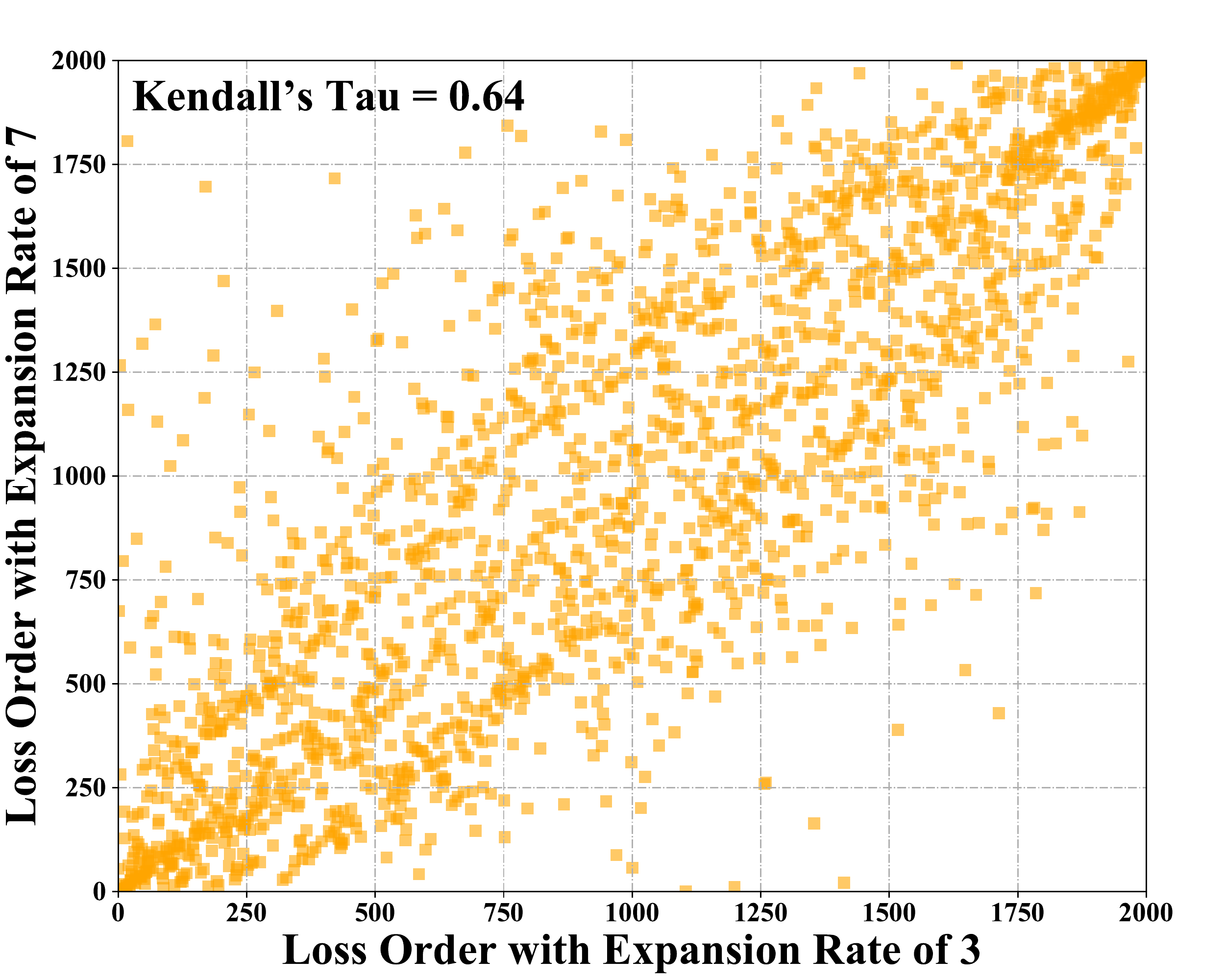}
	\end{subfigure}\hfill
	\begin{subfigure}[t]{0.33\textwidth}
		\includegraphics[width=\textwidth]{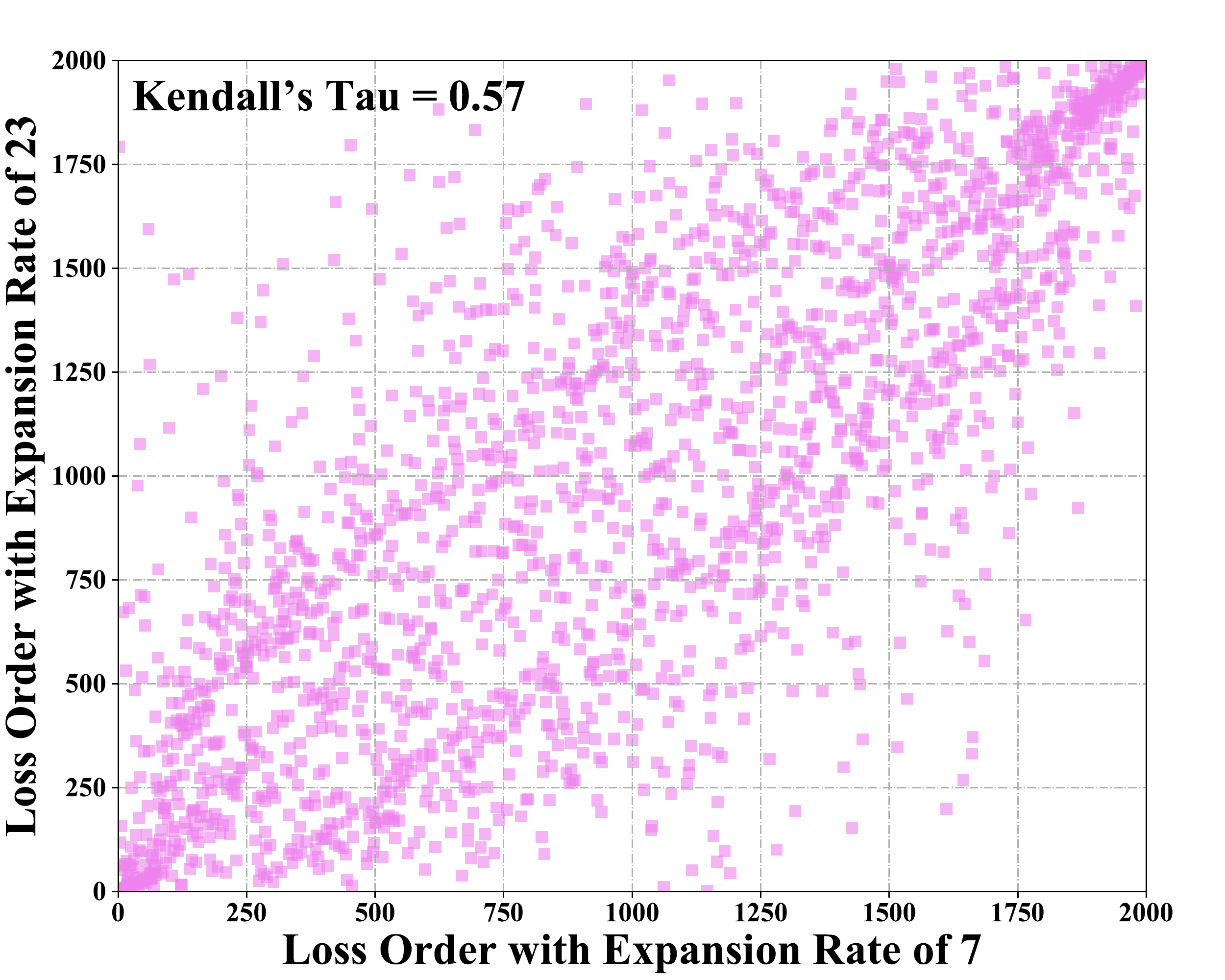}
	\end{subfigure}
	\caption{Expansion rates lead to different 'preferences' on node classification. Taking the prediction of a node at a timestamp as a sample, we order the sample-wise losses in Wikipedia. x and y axes denote different expansion rates. Each point's x value denotes a sample's loss order with the x's expansion rate, while its y value is for the y axis rate. We show the Kerndall's Tau \cite{sen1968estimates} between the x and y values are shown on the left top corner.  \label{fig:exp}}
\end{figure}

TGN may be effective on classifying some nodes at some timestamps, but not others.
This `preferences' varies with different expansion rates, as shown in Fig. \ref{fig:exp}.
In other words, a node at a timestamp can be easy to classify with an expansion rate, but become difficult with another.
We present the Kendall's to evaluate the agreements on the loss orders with different expansion rates in Fig. \ref{fig:exp} as well.
Kendall's Tau is between -1 and 1, and a higher value indicates higher agreements between `preferences' with different expansion rates \cite{sen1968estimates}.
From Fig. \ref{fig:exp}, we observe that the disagreement is larger with a higher gap on the expansion rates.
This meets our expectation, the overlap between the set of sampled neighbors tends to decrease with higher gaps on the expansion rates.
Given the various `preferences' induced by expansion rates, finding the suitable expansion rates for different nodes at any time can further improve TGNs.

\end{document}